# Perovskites for Solar and Thermal Energy Harvesting: State of the Art Technologies, Current Scenario and Future Directions

Richa Pandey[a,b,#], Gaurav Vats[c,#,*], Jae Yun[d], Chris R Bowen[e], Anita W. Y. Ho-Baillie[d] and Jan Seidel[c]

[a]*Centre of Excellence in Nanoelectronics, Indian Institute of Technology Bombay, Powai 400076, India*

[b]*Centre for Research in Nanotechnology and Science, Indian Institute of Technology Bombay, Powai 400076, India*

[c]*School of Materials Science and Engineering, University of New South Wales, Sydney 2052, Australia*

[d]*Australian Centre for Advanced Photovoltaics, School of Photovoltaic and Renewable Energy Engineering*

[e]*Materials Research Centre, Department of Mechanical Engineering, University of Bath, Bath BA2 7AY, United Kingdom*

*Email: er.gauravvats17@gmail.com, g.vats@student.unsw.edu.au; Phone: +61 481354339

[#]Authors contributed equally.

**Abstract:**

Solar energy is anticipated to be the most viable source of sustainable green energy. Perovskites have gained significant research attention in recent years as a solar energy harvesting material due to their desirable photovoltaic enabling properties. The potential strategies for a more effective use of these materials can involve multiple energy conversion mechanisms through a single device or employing materials where a solar or thermal input provides multiple electrical outputs to enhance the overall energy harvesting capability. In this context, the present review focuses on perovskites, including both organic halide perovskites and inorganic oxide perovskites, due to their proven properties as photovoltaic materials and their intriguing potential for additional functionality, such as ferroelectricity. Ferroelectrics are a special class of perovskites that have been studied in detail for photoferroic, pyroelectric and thermoelectric effects and energy storage, which we briefly review here. Furthermore, the possibilities of simultaneously tuning these mechanisms in perovskite materials for multiple energy conversion mechanisms and storage for ultra-high density capacitor and battery applications is also examined in order to attain a better understanding and to present novel opportunities. An understanding of all these mechanisms and device prospects will inspire and inform the selection of appropriate materials and potential novel designs so that the available solar and thermal resource could be utilized in a more effective manner. This review will not only help in selecting an appropriate material from the existing pool of perovskite materials, but will also provide an outlook and assistance to researchers in developing new material systems.

## 1. Introduction

The Kyoto Protocol initiative to reduce carbon emission has endorsed solar energy as the most viable source of sustainable green energy[1, 2]. In this context, solar cells have been deployed faster than anticipated[3] and the solar cell market is expected to exceed a 100 billion USD milestone by 2024; as highlighted in competitive market share and forecast report (2016-2024). In addition, the International Energy Agency (IEA) has anticipated an annual investment of 225 billion USD to achieve power generation levels of 124-200 GW per year using photovoltaic cells and an installed capacity of 4600 GW by 2050 in order to avoid the emission of four gigatonnes of $CO_2$ annually and restrict the mean global temperature rise to $2^0$C, rather than a predicted $6^0$C[1]. Major efforts have been dedicated to reduce the cost and enhance the efficiency of photovoltaics. In this context, novel materials are constantly being explored. Among these materials perovskites have gained significant research interest in recent years because of their low cost and ease of production via soft chemistry[4-9]. *Perovskites* are materials with a $ABX_3$ type structure where cation 'A' occupies the corner positions of the unit cell, cation 'B' is situated at the center of the cell, and anion 'X' is located at the faces of the unit cell; see Figure 1(a) and (b). Perovskites can be classified based on their band gaps as conductors and insulators/dielectrics. Dielectrics with a band gap less than 3eV are termed *semiconductors*. These can also be classified as centrosymmetric, asymmetric and non-centrosymmetric based on their symmetry.

**<Figure 1>**

**Figure 1:** (a) Perovskite structure with symmetric and (b) non-centrosymmetric arrangements. (b)-(c) tuning of the degree of non-centrosymmetry by means of an external stimulus, where *E* is electric field, $T$ is temperature, $hv$ is the photon energy ($h$ is the Planck's constant and $v$ is the frequency of the incident light) and $\sigma$ is the stress.

Figure 2 indicates the classification scheme for perovskites and highlights the domain of interest of this Perspective Paper. The figure indicates that a perovskite can be any material with $ABX_3$ type structure, while ferroelectrics are restricted to the non-centrosymmetric dielectric or semi-conducting materials that possess a spontaneous polarization which can be fully switched by application of an external electric field, stress, thermal fluctuation or light, as illustrated in Figure 1 (b)-(c). The phenomena of achieving a switchable polarization by means of thermal fluctuations is known as the 'pyroelectric effect' while switching of polarization by exposure of light or stress is termed as 'photoferroic effect' and 'piezoelectric effect' respectively. It is to be noted that all ferroelectrics are both pyroelectric and piezoelectric in nature, while the reverse is not true. Figure 3 provides an insight into the relationship between these materials and suggests that a single material could have the desired multiple functionalities which could be utilized simultaneously. Thus, it will be of interest to explore the possibilities of simultaneously harnessing energy from different sources and distinct mechanisms using a single perovskite material, since hybrid perovskites have already been established for photovoltaic[4-9] applications and current research is focused on understanding their behavior on the basis of their crystal structure so that better materials can be developed and designed[10-12].

**<Figure 2>**

**Figure 2:** Classification of perovskites for thermal energy harvesting based on their band gap and crystal symmetry.

In addition, ferroelectrics, a special class of perovskites, are already well known for piezoelectric, pyroelectric and thermal energy conversion systems[13-39], and are also gaining interest as photovoltaic materials[40-51]. Intriguingly, both hybrid perovskites and ferroelectrics are also being investigated for thermoelectric applications[52, 53]. Therefore, the possibilities of utilizing this class of materials to provide multiple functionalities for a more desirable energy output is worthy of consideration. Often, these approaches are considered entirely different branches of research, however considering them simultaneously and holistically can provide several new opportunities. This requires a basic understanding of concepts, mechanisms, corresponding material properties and the underlying physics involved with these different effects. In this context, this perspective aims to provide an understanding of these phenomena as well as state-of-the-art research to motivate researchers from distinct backgrounds and provide pathways to develop better materials systems and devices. The article begins with the basics of solar cells and leads to the emergence of perovskite solar cells. It is then followed by a discussion of photoferroics, pyroelectric energy harvesting, thermo-electric energy conversion and energy storage using supercapacitors and batteries. Finally, the possibilities of coupled mechanisms/devices and future prospects are discussed.

**<Figure 3>**

**Figure 3:** Relationships between perovskite, piezoelectric, pyroelectric and ferroelectric materials

## 2. The Evolution of Solar Cells

Silicon solar cells currently make up for 93% of photovoltaic products of which more than 95% are based on a *p-n* junction architecture[54]. A silicon *p-n* junction solar cell produces power by absorbing light to generate electron-hole pairs followed by the separation of charge carriers by the *p-n* junction and the collection of the electrons (in the *n*-type material as the majority carrier) and holes (in the *p*-type material as the majority carrier) by the electrodes. Figure 4 demonstrates the working principle of a simple *p-n* junction solar cell. The power conversion efficiency ($\eta$) of these cells is expressed as the ratio of output electrical power ($P_{out}$) to the input solar energy ($P_{in}$) absorbed. From the short circuit, current density ($J_{sc}$) and the open circuit voltage ($V_{oc}$), the efficiency can be calculated[55, 56]:

$$\eta = \frac{P_{out}}{P_{in}} = \frac{J_{sc}V_{oc}FF}{P_{in}};\ \mathrm{FF} = \frac{P_{out}}{J_{sc}V_{oc}} \quad (1)$$

where, FF is the *fill factor* and is defined as the ratio of the maximum obtainable power to the product of short circuit current density ($J_{sc}$) and the open circuit voltage ($V_{oc}$) and is limited by parasitic losses such as carrier recombination at the surface and within the bulk, series and shunt resistances. The timeline for development of solar cells can generally be distinguished by the first, second and third generation of solar cells. First-generation solar cells are made of crystalline silicon which dominates the current market. Second-generation cells are typically thin-film photovoltaic cells and despite being an attractive alternative, they come at the expense of a reduced efficiency and contribute to only 7% of the market in 2015. Third-generation of solar cells employ more futuristic concepts and materials including those that utilize electrostatically bound electrons and holes known as *excitons*. In contrast to the directly generated electron-hole pairs in conventional *p-n* junction solar cells these systems have low binding energy and hence induce more current by utilizing photons with comparatively less energy[57]. Third-generation solar cells can further be sub-categorized in organic photovoltaic cells with planar interfaces (OPCPI), organic photovoltaic heterojunction cells (OPHC) and dye-sensitized solar cells (DSSC). OPCPI cells employ an organic polymer that is sandwiched between two metal electrodes with different work functions to confine excitons for light absorption and current generation[57]. These cells have low quantum efficiency, which was later overcome by the development of OPHC cells. Unlike the single polymer layer in OPCPI devices, OPHC cells have two organic layers with a different ionization energy[58] which provides an additional electric potential across the heterostructure and aids in breaking excitons. However, the typical diffusion length of excitons in organic materials is small (on the order of ~10 nm), while a 100 nm absorber thickness is required to produce a sufficient number of excitons. Due to this difference in diffusion length and absorber thickness most of the excitons disappear before reaching the heterojunction and hence merely provide a small contribution towards current generation[58]. This deficiency has encouraged the development of dispersed organic photovoltaic cells and the concept was extensively used in DSSC.

**<Figure 4>**

**Figure 4:** Working principle of a basic *p-n* junction solar cell.

DSSCs have gained popularity as a potential low-cost photovoltaic alternative [4, 6, 59]. The advantage of these cells is that they employ different materials for light absorption and electron and hole transportation. This not only makes these solar cells unique, but it also provides opportunities to tune the performance of these cells by developing new cell materials, cell designs and new cell architectures.

**<Figure 5>**

**Figure 5:** Energy level and device operation of a DSSC. Reproduced from Ref: [6], © 2012 Macmillan Publishers Limited.

The materials for such cells typically consist of wide band gap mesoporous semiconductors of high surface area, such as $TiO_2$, ZnO or $SnO_2$, which are sensitized with a nano-crystalline dye and anchored within a hole-conducting electrolyte or within a hole-transport material (HTM). The system is then sandwiched between two electrodes, one of which is transparent, which is further encapsulated with a glass layer. On exposure to sunlight, the dye absorbs light and a photoexcited electron transfers to the conduction band of the semiconductor and is finally carried to one of the electrodes[60]. Typically, a redox couple transports the positive charge to the other electrode by reducing the oxidized dye to its neutral state[6, 61]. Figure 5 shows the energy level diagram and device operation principle of a typical DSSC. where initially the sensitizing dye absorbs a photon of energy, *hv*, and consequently an electron is injected into the conduction band of the metal oxide; in this case titania. Thereafter, the electron travels to the front electrode (not shown). In addition, the oxidized dye is reduced by the electrolyte, which is regenerated at the counter-electrode to complete the circuit (not shown).

The first DSSC was introduced by O'Regan and Grätzel in 1991[59] for which the overall light-to-electric energy conversion yield was 7.1-7.9% in simulated solar light and 12% in diffuse daylight. Initial complications associated with electrolyte leakage in this cell were later resolved by substituting the liquid electrolyte with a solid hole conductor [62, 63], and such cells are known as solid-state DSSC (ssDSSC). Further progress in the field of DSSCs realized the significance of using ruthenium based organometallic complex sensitizers[64, 65] and an iodide/tri-iodide redox couple[10,11]. The iodide/tri-iodide redox couple provided reduced recombination kinetics which led to longer electron lifetimes of up to 1 s [66-68]. Unfortunately, this system is highly corrosive which has an impact on any metal interconnects and sealants in manufactured devices[6]. These developments led to an improved understanding and to a large increase in the efficiency of DSSCs in the late 1990s, which approached approximately 11.5% in 2006 [69-71].

## 3. The Emergence of Perovskite Solar Cells

Perovskite solar cells have emerged as an advancement of the DSSC[7-9, 72]. Previous parallel research efforts on organic-inorganic hybrids for light emitting diodes (LED) and transistors[73, 74] were anticipated to be applied to solar cell applications by Mitzi and co-workers[74]. However, this did not gain significant attention, which is possibly due to environmental issues associated with the use of lead (Pb) and concerns regarding the robustness of tin (Sn) based perovskites[8]. However, the first reports on the photovoltaic response of organometallic perovskites are attributed to Miyasaka's group with documented efficiencies of 2.2% (2006) and 3.8% (2009) in $CH_3NH_3PbBr_3$ and $CH_3NH_3PbI_3$, respectively[75-77]. Thereafter, Park and co-workers (2011) achieved an efficiency of 6.5% by optimizing the perovskite coating solution concentration, post-annealing condition and $TiO_2$ surface modification[78]. Although the performance of their perovskite sensitizers ($(CH_3NH_3)PbI_3$ quantum dots) was better than the standard N719 dye sensitizers, they possessed a poor stability and dissolved in the electrolyte under continuous irradiation after only 10 minutes[78]. In order to address their stability, the Park and Grätzel groups combined efforts for the replacement of the electrolyte by a solid-state hole transport material, namely spiro-MeOTAD (2,2',7,7'-tetrakis(N,N-di-p-methoxyphenylamine)-9,9'-spirobifluorene)[79]. This not only enhanced the stability of the cell, but also improved the efficiency to 9.7%[79]. In the same year (2012), Snaith and colleagues introduced four additional developments along with spiro-MeOTAD and reported an efficiency of 10.9%[80]. Further efforts considered the following aspects (a) the use of mixed halide $CH_3NH_3PbI_{3-x}Cl_x$ to achieve an improved stability and carrier transport in contrast to pure iodide and bromide equivalents[80, 81]; (b) forming an extremely thin absorber (ETA) by coating a thin perovskite layer onto $TiO_2$; (c) replacing the conducting nano-porous $TiO_2$ with a non-conducting $Al_2O_3$ network; (d) utilizing ambipolar transport by aid of planar cells without any scaffolding[8, 80]. This work exploited the fact that perovskites are capable of transporting both electron and holes to the cell terminals, rather than merely working as sensitizers[8]. In 2013, Seok and Grätzel introduced a polymeric hole conductor (poly-triarylamine) with a three-dimensional nanocomposite of mesoporous-$TiO_2$ and $CH_3NH_3PbI_3$ perovskite as the light harvester[82]. This led to a power conversion efficiency of 12.0%, with a substantial improvement in the open circuit voltage and fill factor of the cell[82]. Seok and co-workers replaced $CH_3NH_3PbI_3$ with $CH_3NH_3PbI_{3-x}Br_x$ and raised the efficiency further to 12.3%[83] which was followed by attempts to attain a better morphology of the perovskite layer for improved efficiency. This included the use of sequential deposition in which the initial $PbI_2$ solution was introduced into a nanoporous titanium dioxide film and later exposed to $CH_3NH_3I$ for transformation into $CH_3NH_3PbI_3$ perovskite[84]. Similarly, a mixed halide ($CH_3NH_3PbI_{3-x}Cl_x$) was deposited by a two-source thermal evaporation and an efficiency of 15.4% was achieved[85]. By the end of 2013, the efficiency approached 16.2% using a mixed halide $CH_3NH_3PbI_{3-x}Br_x$ (10-15% Br) and a poly-triarylamine HTM[86]. In 2014, a confirmed efficiency of 17.9%[8] was reported by mixing the lower bandgap $CH(NH_2)_2PbI_3$ with the $CH_3NH_3PbBr_3$ as the light-harvesting layer[87], there was also an unconfirmed efficiency of 19.3%[8, 88]. Zhou *et. al*. fabricated $CH_3NH_3PbI_3$ perovskite on doped $TiO_2$ with an yttrium and modified indium tin oxide cathode with polyethylenimine ethoxylated to reduce the contact barrier[89]. An efficiency of 20.1% was independently confirmed in late 2014, as demonstrated by Seok and co-workers[90]. In this work, high-quality $FAPbI_3$ films were fabricated by direct intramolecular exchange of dimethylsulfoxide (DMSO) molecules intercalated in $PbI_2$ with formamidinium iodide. To date, the highest reported efficiency for a perovskite solar cells is at 22.1%, demonstrated by the team at the Korea Research Institute of Chemical Technology (KRICT) and Ulsan National Institute of Science and Technology (UNIST)[91]. Recent work further extended the efficiency to 23.6% by employing a two-terminal tandem configuration using an infrared-tuned silicon heterojunction bottom cell and a caesium formamidinium lead halide perovskite top cell [92]. Figure 6 summarizes the progress to date and provides an insight into the existing DSSC and perovskite solar cells.

**<Figure 6>**

**Figure 6:** Timeline for the evolution of perovskite solar cells; beginning with the discovery of (a) a DSSC which was comprised of a liquid electrolyte and dye sensitized mesoscopic $TiO_2$[59]. Electrolyte leakage was eliminated by introducing a solid state organic p-type hole conductor ((b) ssDSSC)[62, 63]. Later, in (c) extremely thin absorber (ETA) cells replaced the dye with an extremely thin absorber semiconductor layer[93, 94]. The efficiency was further improved in (d) meso-superstructured solar cell (MSSC) by using a perovskite layer and a porous insulating scaffold instead of an ETA and $TiO_2$ respectively[80]. It is suggested that the efficiency is likely to improve by employing (e) porous perovskite *p-n* heterojunctions[7] and (f) *p-i-n* thin film perovskite solar cells in which a thin perovskite film is sandwiched between *p*-type and *n*-type charge-extracting contacts [7, 95, 96].

At this stage, it is important to note that unlike DSSCs, perovskite solar cells utilize a single perovskite layer for both light absorption and electron-hole transportation. However, a barrier to commercial deployment of these materials is their poor stability and the toxicity of lead in the most efficient perovskite cells [97, 98]. In this context, Giustino and Snaith have provided an insight on various possible lead-free alternatives, which are shown in Table 1 of reference [9]. However, the continued investigation and search for improved materials is warranted as the efficiency achieved using lead-free counterparts remains relatively low. Recently, Wang *et. al.* reported a potential restriction on further progress in iodide based perovskite cells[99]. They revealed that iodide based perovskites produced gaseous iodine ($I_2$) during operation, which has a high vapor pressure and therefore permeates through the perovskite layer and results in degradation of the material[99]. However, they did not completely rule out the possibility of using iodide perovskites for solar cells, but strongly advocated the need to develop new stable perovskites. Recently, Shin *et. al.* employed inorganic perovskite, La doped $BaSnO_3$, as an electron transport layer instead of a conventional $TiO_2$ and achieved a remarkable efficiency of 21.2% and photostability[100]. It retains 93% of its initial performance after 1,000 hours of exposure to sunlight.[100] In addition, parallel research is ongoing to understand the existence and possibility of ferroelectricity[101-104] in such materials which is often overlooked and has also become a question of debate in recent years[105-109]. Interestingly, inorganic ferroelectric materials are also being explored for solar energy harvesting and the effect is known as photo-ferroelectric/photoferroic or ferroelectric photovoltaic[42-45, 110-114].

## 4. Ferroelectric Photovoltaics and Photoferroics

Polar materials are being extensively studied as a potential alternative to semiconductor-based materials for photovoltaic applications[40, 41, 46, 49, 115-122]. The dipole moment in these materials due to exposure to light, heat or by inducing an internal field and interface band bending when an external voltage is applied to the system facilitates the generation of charge carriers at the material-electrode interface. Among these materials, ferroelectrics have additional advantages such as piezo- and pyro-electricity, as described in Section 1. Research in this area started with the discovery of the generation of a photocurrent in paraelectric $BaTiO_3$[123], this was followed by the detection of above band gap photovoltages in cadmium telluride[124-126] and zinc sulfide thin films[127]. Later, in the early 1970s, a *bulk* or *anomalous photoferroic/photovoltaic* effect was discovered in non-centrosymmetric crystals; this is also known as the *galvanic effect* or *non-linear photonics*[43, 51]. This effect in ferroelectric and multi-ferroic materials refers to the phenomena of obtaining a steady state current in short circuit condition or a high output photovoltage in the open circuit condition in the direction of polarization of the materials on exposure to continuous illumination[42, 47]. Initial investigations focused on bulk materials, but it was later observed and studied at the nanoscale, which became a reason for the effect to be described as a '*bulk*' photovoltaic effect (BPVE); while the term '*anomalous*' photovoltaic effect (APVE) was used due to experimental observations of photovoltages $10^3$ to $10^4$ times higher at open circuit in contrast to the band gap of the material[42, 46]. Not all ferroelectrics exhibit an APVE, as it is dependent on the polarization magnitude[128], direction of polarization[41, 48, 129], light intensity[130], electrode spacing[131, 132], electrical conductivity[46] and the crystallography of the material[131, 133, 134], in addition to the nature of domain walls[40, 41] and material/electrode interfaces[115]. Its dependence on so many factors often leads to difficulty in reproducing the APVE, even in the same material[112, 113]. Therefore, several models have been proposed to explain the distinct type and nature of photoferroic effects. These include Schottky-junction effects, depolarization field effects and interface and domain wall effects[40, 41], which are now described below.

### 4.1. Bulk Photovoltaic Effect (BPVE)

The first model of BPVE was proposed by Glass et. al.[135] which was based on the asymmetry in materials. In recent years, Rappe and co-workers have developed theories based on shift currents[136-140]. Consequently, BPVE and APVE could now be explained at the microscopic level using *ballistic* and *shift* mechanism models[47, 49]. The *ballistic model* for isotropic and anisotrpic materials in centrosymmetric (in the general case for p-n junction solar cells) and non-centrosymmetric crystals (in the general case for ferroelectric solar cells and asymmetric hybrid perovskite solar cells) is explained using Figure 7 (a) and (b), respectively[50]. This manifests itself in that on exposure to an appropriate illumination the thermalized/hot carrier from the valence band excites to the conduction band, thereby leading to the generation of a photocurrent. However, the presence of asymmetry or non-centrosymmetry in crystals leads to a disparity in a momentum distribution of the carriers in the conduction band. The carrier then loses its energy and settles at the bottom of the conduction band by undergoing a band-band transition or a shift by distance $l_o$ so as to equilibrate the asymmetric momentum and hence generating an additional photocurrent leading to BPVE[49, 50].

< **Figure 7**>

**Figure 7:** (a) Isotropic and (b) anisotropic non-equilibrium carriers' momentum distribution in centrosymmetric (general case for p-n junction solar cells) and non-centrosymmetric (general case for ferroelectric solar cells and asymmetric hybrid perovskite solar cells) corresponding to the classical and bulk photovoltaic effects, respectively. Adapted from references[47, 50]

In contrast, the *shift current* mechanism has a quantum-mechanical nature and the behavior of the thermalized carriers is governed by coherent excitations, rather than inelastic scattering, which allows for the net current flow from the asymmetry of the potential[136, 140, 141]. The same mechanism is also supported by experimentally verified first principle studies on $BaTiO_3$ and $PbTiO_3$ ferroelectrics[136, 137], multiferroic $BiFeO_3$ [137]and hybrid halide perovskites $CH_3NH_3PbI_3$ and $CH_3NH_3PbI_{3}$-xClx [138]. It is also suggested that the material itself can act as a current source[97,101,104] and the effect is also dependent on electronic structure and bonding interactions[136]. The total photocurrent ($J$) of a ferroelectric material in the closed circuit condition can be given as the sum of steady current density ($J_{sc}$) generated due to illumination and the contribution of dark- ($\sigma_d$) and photo- ($\sigma_{ph}$) conductivities:

$$J = J_{sc} + (\sigma_d + \sigma_{ph})E \quad \text{or} \quad J = J_{sc} + (\sigma_d + \sigma_{ph})\frac{V}{d} \qquad (2)$$

where $E$ is the internal electric field developed between electrodes separated by distance $d$, and $V$ is the applied voltage. In the open circuit condition, the total current ($J$) will vanish and hence the open circuit voltage ($V_{oc}$) is given as:

$$V_{oc} = \frac{J_{sc}d}{(\sigma_d + \sigma_{ph})} \qquad (3)$$

The above expression suggests that $V_{oc}$ will be anomalous if illumination leads to a significant rise in steady current density ($J_{sc}$). In addition, it is to be noted that the photoconductivity ($\sigma_{ph}$) is also dependent on light intensity[49, 51, 135, 142-144]. If the rise in $\sigma_{ph}$ during illumination is of the order of the rise in $J_{sc}$, then it will cancel out the influence of a rise in $J_{sc}$ and cause the material to exhibit a constant or linear photoferroic effect. However, if there is a condition where $J_{sc} >> \sigma_{ph} >> \sigma_d$ then the effect will be 'anomalous' since $V_{oc}$ in this case will increase abruptly. In addition, if $\sigma_{ph}+\sigma_d$ is insensitive to the light intensity, or the change in $\sigma_{ph}$ is small in comparison to the magnitude of $\sigma_{d,}$ then this condition will also contribute towards the BPVE/APVE. In general, for most of the ferroelectric materials the case of $J_{sc}>>\sigma_{ph}>>\sigma_d$ exists and hence the $V_{oc}$ for BPVE can be simplified to $\sim J_{sc}/\sigma_{ph}$. This can be further explained by substituting the following ($\mu$: mobility of non-equilibrium charge carrier; $\alpha$: absorption coefficient; $\tau$: life time of non-equilibrium charge carrier; $\xi$: measure of exciton; $l_0$: mean free path; asymmetry; $hv$: photon energy; $\varphi$: quantum yield; $q$: positive elementary charge; $\Delta n$: excess charge carrier concentration; $\phi_0$: photon flux density)[50, 145]:

$$J_{sc}=ql_0\xi\varphi\alpha \tag{4}$$

$$\sigma_{ph}=q(\mu_n+\mu_{ph})\Delta n= q(\mu_n+\mu_{ph})\varphi\alpha\tau\phi_0 \tag{5}$$

Hence, the efficiency of the BPVE/APVE can be calculated by using equation 1 and expressed as[50, 145]:

$$\eta = \frac{J_{sc}^2}{\alpha I \sigma_{ph}} = \frac{J_{sc}^2}{\alpha I_0 (\mu\tau)_{ph}} \text{ or } = \frac{q\alpha\varphi(l_0\xi)^2 d}{4(\mu_n+\mu_p)\tau h\nu} \text{ with an assumed FF of 25\%} \tag{6}$$

Interestingly, the BPVE/APVE was first explained in ferroelectrics using the aforementioned models. However, the first evidence of the BPVE/APVE effect was reported in paraelectric $BaTiO_3$[123], non-ferroelectric cadmium telluride[124-126] and zinc sulfide thin films[127]. It was suggested that this is due to the formation of surface space-charge layers[123] or stacking faults that produced a cumulative internal depolarization field[127]. Simultaneously, the same reasons were also thought to affect the process of domain nucleation[123].

### 4.2. Depolarization Field Driven Ferroelectric Photovoltaic Effect

Ferroelectric materials possess a spontaneous polarization, i.e. electric dipoles are formed inside the material. Ideally, if the ferroelectric is sandwiched between electrodes with the same work-function then the built-in voltage due to the presence of dipoles must be balanced by the presence of charges at the electrodes. However, in practice the free charges at the electrodes are not able to completely cancel the space and polarization charges which gives rise to internal fields in the opposite direction of polarization[146]. The cumulative internal field developed by these unscreened charges accumulated at the ferroelectric-metal interface is known as a *depolarization field* [145, 147]. It has been shown that the depolarization field is capable of changing the overall magnitude of the polarization, transition temperature, coercive field and the order of the phase transitions [148, 149]. Interestingly, the polarization filed is dependent on the material as well as electrode thickness and the area of contact [148, 149].; it is negligible for a large inter-electrode distance in bulk ferroelectrics but is likely to increase with a reduction in inter-electrode distance, as in thin films[131, 146, 148, 149]. This eventually makes it a governing factor for the photo-ferroelectric effect in thin films as they significantly influence both the screening of spontaneous polarization and the separation of the photo-generated charge carriers[121, 150-152]. In this context, Pintilie and Alexe postulated that the polarization bound charge is not located at the electrode but is slightly away at an atomic distance $\delta$ from the electrode and hence results in the formation of surface dipole layers that lead to a modification in the surface injection barriers [153]. Furthermore, this built-in voltage ($V_{bi}$) due to the difference in work function of the electrodes is modified by the surface dipole layers and is given as[145]:

$$V'_{bi} = V_{bi} + \frac{P_s \delta}{\varepsilon_0 \varepsilon_r} \tag{7}$$

The depolarization field for a metal contact with different dielectric constants ($\varepsilon_{e1}$ and $\varepsilon_{e2}$) and screening lengths ($l_{s1}$ and $l_{s2}$) for a dielectric constant ($\varepsilon_F$) can be written as[154, 155]:

$$F_{dp} = \frac{P}{\varepsilon_0 \varepsilon_F} \frac{\left(\frac{l_{s1}}{\varepsilon_{e1}} + \frac{l_{s2}}{\varepsilon_{e2}}\right)}{\left(\frac{l_{s1}}{\varepsilon_{e1}} + \frac{l_{s2}}{\varepsilon_{e2}}\right) + d} \tag{8}$$

The impact of the depolarization field on photovoltaic performance of thin films has been verified in several reports and it is believed that ultra-high thin films (films with thickness of a few nanometers) with high dielectric constant electrodes can aid in achieving high photovoltaic efficiencies [130, 131, 133, 146, 148-152, 156-158]. Furthermore, it has been illustrated that it is possible to control the transportation characteristics by controlling the direction of polarization[159]. The control over transportation characteristics is also related to the Schottky barrier, which is another important mechanism for ferroelectric photovoltaics and will be discussed in the next section.

### 4.3. Ferroelectric Schottky-Junction Effect

A Schottky barrier is formed at a ferroelectric-metal electrode interface due to the difference in work-functions, which leads to the development of a local electric field. On illumination, this built-in field drives the photocurrents by band bending at the interface[160]. Therefore, the barrier height and the depth of the depletion region plays an important role in the generation of photovoltages[160]; which is by the constraints of the material band-gap and work-function of the electrode [113, 145, 147]. Due to this reason the effect is probably less well studied in contrast to the bulk photovoltaic effect (BPVE). The overall barrier height can be enhanced by sandwiching a ferroelectric semiconductor between electrodes of different materials with large difference in work function[115, 159, 161-168]. This was first demonstrated by Blom *et. al.* in 1994[169] who sandwiched a ferroelectric $PbTiO_3$ film between a Schottky contact (Au) and an Ohmic bottom electrode ($La_{0.5}Sr_{0.5}CoO_3$) and showed that the Schottky barrier can be reduced by switching the polarization in the direction of the ferroelectric polarization. Another popular mechanism of tuning the barrier height and the width of depletion region is to tune charged defects[163, 170]. In this context, the most common defect, namely oxygen vacancies, have been studied in hybrid halide perovskites[171, 172] and ferroelectrics[115, 170]. However, the photovoltaic effect obtained using this mechanism is not very stable as the poled state of oxygen vacancies usually becomes unstable on removal of the electric field over time. However, it remains very useful in distinguishing between the depolarization and Schottky junction based photovoltaic mechanism as it is possible to have switchable diode-like rectifying behavior using the Schottky junction effect[163, 166, 169] but not with the depolarization field[115]. Interestingly, the Schottky junction effect is independent of the direction of polarization and, therefore, it can be used to distinguish it from BPVE[173]. The understanding developed to date can aid in tuning the Schottky junction effect to support BPVE and achieve a combined photovoltaic response.

### 4.4. Interface and Domain Wall Effects

Domain walls in complex oxides have been the focus of intense research over recent years. The fact that domain walls can be electrically conducting opens new pathways for a number of possible applications.[119, 174-177] Recently anomalous photovoltaic effects related to domain walls in ferroelectric materials have been reported.[40, 175, 178-184] Interestingly, electric-field control over domain structure allows the photovoltaic effect to be reversed in polarity or even to be turned off in such materials. The band structure and local bandgap of domain walls in ferroelectrics have also been studied.[185-188] In addition, photo-induced electrochemical effects at domain walls are a further interesting route in applications in water splitting[189] or for domain wall decoration (see[190] and references therein).

The spatial and temporal evolution of photoinduced charge generation and carrier separation in heteroepitaxial $BiFeO_3$ thin films was measured with Kelvin probe and piezoresponse force microscopy.[191] Contributions from the self-poled and ferroelectric polarization charge were identified from the time evolution of the correlated surface potential and ferroelectric polarization in both, films as-grown and after poling, and at different stages and intensities of optical illumination. Variations in the surface potential with bias voltage, switching history, and illumination intensity were investigated. It was shown that both bulk ferroelectric photovoltaic and the domain wall offset potential mechanisms contribute to the photogenerated charge. Polycrystalline[150], 2-D interfaces,[192-195] and 1-D ferroelectric nanostructures have also been explored for enhanced photovoltaic responses[196, 197,] in addition to nanoscale enhancements of ferroelectric photovoltaic effects at metal nano-tips.[116]

### 4.5 An Overview of Ferroelectric Photovoltaic Materials

Table 1 provides an insight into the photo-ferroelectric response of selected ferroelectric materials and their corresponding mechanism. In addition to these attempts and a history of over 40 years of study, there remains much to be learned before a ferroelectric photovoltaic device reaches any niche applications market. Although it is understood that non-centrosymmetry and polarization plays a crucial role in the bulk and depolarization field driven photovoltaic effect, the dynamics of the process remains unclear. The understanding of how exactly ferroelectricity helps in achieving an enhanced photo-response is an open question [168]. Moreover, from the photovoltaics view-point, there is a strong requirement of developing new ferroelectrics with a narrow bandgap and improved conductivity. Thereafter, domain wall engineering and a controlled polarization can aid in raising the photovoltaic efficiency to new levels. In addition, there is a need to systematically understand charge carrier dynamics, such as mobility and diffusion length. Beyond this, other important considerations are the presence of piezo- and pyro-electricity in ferroelectrics. The photovoltaic response of ferroelectrics could be tuned under stress or hydrostatic pressure. Recently, Wang *et. al.* made an analogous attempt on hybrid perovskites and realized that it is possible to tune both the structural and optical properties by applying a pressure on the material[198]. However, this needs special arrangement, but the pyroelectric effect could be exploited in parallel to photovoltaic effect to enhance the overall energy conversion.

## 5. Pyroelectric Effect

The effect of generating an electric charge due to changes in remnant and/or saturation polarization as a result of thermal fluctuations is known as the *pyroelectric effect*[31, 37, 199, 200]. Figure 8 shows the schematic of time dependent thermal fluctuations that lead to a displacement of the central ion in a non-centrosymmetric perovskite and results in a change in output voltage. This can be used to supply an electrical current by using a resistive load. The change in polarization $(\Delta P_i)$ with temperature change $(\Delta T)$ is given as[201]:

$$\Delta P_i = p\Delta T \tag{9}$$

where, $p$ is the pyroelectric coefficient perpendicular to direction of the electrodes (i.e. in the polarization direction). Further, for a given surface area $A$, the induced short circuit current ($I_P$) for a given rate of temperature change $(dT/dt)$ is[201, 202]

$$I_P = Ap\frac{dT}{dt} \tag{10}$$

In order to characterize enhanced pyroelectric energy conversion various figures of merit (FOMs) have been developed for the selection of appropriate pyroelectric materials depending on the thermal system and electrical circuits employed[202]. The first pyroelectric FOMs were based on the maximum current or voltage for applications related to thermal detectors[203-205].

< **Figure 8**>

**Figure 8:** Schematic presentation of pyroelectric effect where time dependent thermal fluctuations cause the displacement of the central atom in a non-centrosymmetric perovskite and results in an output voltage.

To achieve a high voltage responsivity ($F_v$), the figure of merit[203] to maximize the voltage for a given thermal input is given by Eqn. 11,

$$F_i = \frac{p}{c_E} = \frac{p}{\rho . c_p} \tag{12}$$

In the case of a pyroelectric detector dominated by Johnson noise, the detector figure of merit is[202],

$$F_D = \frac{p}{c_E\sqrt{\varepsilon_{33}^{\sigma} tan\delta}} \tag{13}$$

Eqns. 11 to 13 are FOMs that have been used to select materials for thermal detection, however for energy-harvesting the generated power is a criterion, along with the efficiency of the conversion of thermal energy to electrical energy. For energy harvesting applications, pyroelectric FOMs have been proposed[36, 206] and an electro-thermal coupling factor estimates the effectiveness of thermal harvesting[36]:

$$k^2 = \frac{p^2.T_{hot}}{c_E.\varepsilon_{33}^{\sigma}} = \frac{p^2.T_{hot}}{\rho.c_p.\varepsilon_{33}^{\sigma}} \quad (14)$$

where $T_{hot}$ is the maximum working temperature. An energy-harvesting FOM, $F_E$, has also been proposed[206]:

$$F_E = \frac{p^2}{\varepsilon_{33}^{\sigma}} \quad (15)$$

Eqn. 15 has been used for materials selection and materials design[29, 207-210] for pyroelectric harvesting applications. Compared to the voltage ($F_v$) and current ($F_i$) responsivities, $F_E$ does not consider the material heat capacity. Therefore, a modified pyroelectric thermal harvesting figure of merit, $F_E'$, has been derived when the harvesting device is subjected to an incident heat source of specific energy density [211-213], which is given by:

$$F_E' = \frac{p^2}{\varepsilon_{33}^{\sigma}.(c_E)^2} \quad (16)$$

Eqn. 16 indicates that good materials should have a high pyroelectric coefficient to develop a large charge with a temperate change, a low permittivity to develop a large potential difference because of the charge generated and a low volume specific heat to ensure the temperature rise due to the incident power density is large. It may be of interest to also consider losses and develop a figure of merit which includes loss, such as the tan$\delta$ used in $F_D$ (Eqn. 13).

In addition to the figures of merit for pyroelectric energy conversion, it has been suggested that the short circuit current can be enhanced to several orders of magnitude by employing thermal energy conversion cycles. In this context, Mohammadi and Khodayari advocated for the use of an Ericsson cycle [214] knowing that there exist several other cycles based on the mode of operation[36, 39]. These include resistive cycles[36], synchronized electric charge extraction cycles[36, 39] and synchronized switch damping on inductor (SSDI) cycles. Details about these cycles can be found in reference[22]. It has been found that the *Olsen cycle,* a well-known variant of the Ericson cycle, stands out in this regime[215-217].

### 5.1. Olsen Cycle

The Olsen cycle operates under unipolar electric fields, rather than bipolar electric fields used in the conventional Ericson cycle, and therefore has a reduced hysteresis loss and enhanced energy conversion. The energy harvested using the Olsen cycle is not merely a contribution of the pyroelectric effect but also takes advantage of the electrical energy storage capacity of the material as a result of the change in capacitance (permittivity) with temperature. For this reason, the cycle is claimed to be capable of providing an energy density of three orders in magnitude higher than that obtained using the conventional pyroelectric effect [23]. Olsen *et. al.* experimentally verified this claim for a number of well-known compositions[215-222] which has been supported by a number of studies [13-17, 23-25, 27, 209, 223]. This made the Olsen cycle the primary mechanism for pyroelectric energy harvesting. Since the cycle is based on temperature dependent polarization behavior, it is important to note that, in general, the saturation polarization decreases with an increase in temperature and such a response is referred to as a 'thermal fluctuations-1 (TF-1)' behavior while the case where the saturation polarization increases with an increase in temperature, i.e. where the hysteresis loop tends to become linear at low temperatures rather than at high temperatures, is known as 'thermal fluctuations-2 (TF-2)' behavior[17, 20]. Since TF-2 compositions were rarely observed before 2008, the Olsen cycle was initially proposed for the commonly observed TF-1 ferroelectrics. In 2014, Vats *et. al.* proposed a modified version of the Olsen cycle for TF-2 compositions [17] and generalized it for materials that exhibit a change in polarization with temperature fluctuations[17, 20]. Figure 9 (a) and (b) show the working principle of an Olsen cycle for TF-1 and TF-2 composition, respectively.

< **Figure 9**>

**Figure 9:** Working of an Olsen cycle for (a) TF-1 and (b) TF-2 compositions and the area covered by the loop 1-2-3-4 shows the harvested energy.

The modified cycle states that the material should initially be polarized under a unipolar applied electric field at the lower temperature *($T_L$)* and then exposed to a heat source isoelectrically *($E_H$)*. This leads to a polarization change (a decrease for TF-1 and increase for TF-2) that can be simultaneously converted into an electrical output. Subsequently, the material is depolarized under a unipolar applied electric field at a constant higher temperature *($T_H$)* followed by an isoelectric *($E_L$)* cooling step. This again provides an output electrical impulse in the form of harvested electrical energy. Figure 10 (a) and (b) provide a schematic explanation of a typical Olsen cycle for TF-1 and TF-2 materials, respectively. Further, the area enclosed *(1-2-3-4)* by the complete cycle on a corresponding *P-E* curve gives the net harvested output electrical energy density ($N_D$) per liter per unit cycle [24, 221]:

$$N_D = \oint E.\mathrm{d}P \tag{17}$$

Table 2 summaries the performance of selected compositions for pyroelectric energy conversion. It is important to note that for a material to work with a practical Olsen cycle, an arrangement is needed to achieve specialized oscillating heat currents and an external load circuit is required for receiving an electrical output. A variation in the design of such an arrangement can significantly influence the degree of harvested electrical energy density. It is important to note that pyroelectric energy conversion is based on a time dependent thermal fluctuation. At the same time, there exists a mechanism which utilizes a thermal gradient to harvest electrical energy, as discussed in the next section.

## 6. Perovskites for Thermoelectrics

### 6.1. Ferroelectric-Thermoelectrics

Thermoelectric generators are devices which convert temperature differences into electrical energy. The principal phenomenon which underpins this energy conversion are the Seebeck effect (i.e. the conversion of a thermal gradient into electricity) and Peltier effect (i.e. achieving a temperature gradient by passing a current through two junctions).[224-226] The thermoelectric effect have been widely employed for scavenging of waste thermal energy and the efficiency of a material for thermoelectric application is measured by its figure of merit, *zT*: [52]

$$zT = S^2 \sigma T / \kappa \tag{18}$$

Where, *S* is the Seebeck coefficient and is defined as the ratio of the voltage change induced by a temperature change for a material with thermal and electrical conductivities $\kappa$ and $\sigma$ respectively. $S^2\sigma$ is termed the thermoelectric power factor. To achieve a high thermoelectric efficiency the material should have high electronic charge carrier concentration ($\sim 10^{18}$ to $\sim 10^{21} cm^{-3}$) and high electronic conductivity. An increase in carrier concentration will not only increase the electric conductivity but simultaneously enhance thermal conductivity which results in a decrease of the *zT* value and reduced thermoelectric performance of the material. Apart from this, the major roadblocks in the development of materials for this technology include a limited working temperature range, the use of toxic chemicals and high processing costs, while some materials such as chalcogenides and antimonides have issues of oxidation at high temperatures.[52] In the search of potential candidate materials which can overcome the above mentioned issues, oxygen-deficient ferroelectrics with high conductivities have provided a new direction.[227, 228] $SrTiO_3$ based materials have been well studied but their *zT* values are limited to 0.1 due to moderate thermal conductivities.[229-236] In addition, $CaMnO_3$-based systems have a high Seebeck coefficient, low thermal conductivity and tunable resistivity[237, 238], which are ideal properties for improved thermoelectric applications. Theoretical estimations suggests that it is possible to have *zT* values greater than one in $CaMnO_3$.[239] Lee *et. al.* reviewed future directions in ferroelectric based thermoelectricity and stressed the potential of obtaining a high thermoelectric effect in n-type perovskite $BaTiO_3$ and tungsten bronze $(Sr_{1-x}Ba_x)Nb_2O_{6-\delta}$ systems[52] and concluded that improved thermoelectric performance of ferroelectrics could be achieved by a better understanding of the mechanism behind the electronic interactions, defect states and oxygen vacancies. However, from a device perspective it is easy to optimize a pyroelectric (as it is easy to have time dependent temperature change rather than inducing a large thermal gradient over a ferroelectric thin film or pellet) based system instead of a thermal gradient (thermoelectric effect) based system on ferroelectric materials.[33] Consequently, a maximum Carnot efficiency of merely 1.7% was reported using the thermoelectric effect, while the same is found to be 50% for a pyroelectric device.[33]

### 6.2. Hybrid Halide Perovskites for Thermoelectrics

In parallel research efforts on ferroelectric thermoelectrics, there exists a branch of organic thermoelectric materials which have recently gained significant interest among researchers because of their relatively low cost of manufacture, ease of fabrication and possibility of developing flexible thermoelectric modules[240]. He and Galli conducted a pioneering first principle study of $CH_3NH_3AI_3$ (A=Pb and Sn) for thermoelectric applications and realized that it is possible to have a $zT$ in the rage of 1 to 2 by engineering hybrid halide perovskite superlattices.[53] Their study suggests that these perovskites may possess a large carrier mobility due to small carrier effective masses and weak carrier-phonon interaction. These materials have a large Seebeck coefficient and low thermal conductivities which makes them ideal candidates for thermoelectric energy harvesting. However, their electrical conductivity needs to be enhanced and this could be done by chemical or photoinduced doping.[241] Lee *et. al.* have used Density Functional Theory to suggested that $CH_3NH_3AI_3$ has a poor thermoelectric performance but it can be increased to the levels of the existing best thermoelectric counterparts ($Bi_2Te_3$) by electron-doping.[242] The claim was further supported in a theoretical study by Filippetti *et. al.* where they reported the possibilities of achieving a room temperature thermoelectric effect with $zT$ values ranging between 1 to 3.[243] Thereafter, Wang and Lin conducted Molecular Dynamics simulations and provided atomistic insights on ultralow phonon transport over a wide temperature range in these materials.[244] Recently, Zhao *et. al.* conducted a first principle study and suggested that hole-doping optimization could provide better thermoelectric performance over the electron-doped one.[245] In addition, they proposed to tailor the organic cation vacancies for better thermoelectric performance. By taking advantage of tuning the electrical conductivity by optimized doping, these studies postulate that hybrid halide perovskite more suitable for thermoelectric energy conversion in contrast to their ferroelectric counterparts. However, at this stage it is difficult to accurately predict the exact status of these materials as thermoelectric generators due to the lack of experimental confirmation.

## 7. Perovskites for Energy Storage

### 7.1. Ferroelectric Perovskites for Capacitor Applications

Ferroelectric materials and their composites have been studied in detail for ultra-high density capacitor applications[246-252]. When an electric field is applied to a ferroelectric material sandwiched between two electrodes, it is polarized and energy is stored. The maximum energy stored (*U*) in such a capacitor is given by[253]:

$$U = \frac{1}{2}\frac{CV^2}{Volume} = \frac{1}{2}\frac{\varepsilon\varepsilon_0 A}{t}\frac{(E_b t)^2}{Volume} = \frac{1}{2}\frac{\varepsilon\varepsilon_0 A}{t}\frac{(E_b t)^2}{tA} = \frac{1}{2}\varepsilon\varepsilon_0 {E_b}^2 \qquad (19)$$

where, $C$ is the capacitance and $E_b$ is the dielectric breakdown strength of the intervening dielectric layer of thickness $t$ and electrode contact area $A$. Eq 19 suggests that the energy stored in a ferroelectric capacitor is dependent on the dielectric constant ($\varepsilon$) and breakdown strength of the material. Therefore, research has focused on improving both parameters. In this context, ferroelectric-polymer and ferroelectric-glass composites are being explored and have been proven to be good alternatives [248-250]. In addition, the energy stored can also be estimated from the polarization versus electric field behavior of the materials, as illustrated by the shaded area of Figure 10. This clearly indicates that for high energy storage, the material should have a low remnant polarization and high saturation polarization with a low hysteresis, as observed in the case of relaxor ferroelectrics in Figure 10 (b). It is to be noted that in addition to these parameters the material should also possess low leakage current and dielectric losses.[248, 250-252] A more detailed discussion of the state-of-the-art progress can be found in a recent review by Liu *et al.*[246] Although ferroelectrics have a proven potential as a dielectric material in a parallel plate capacitor structure, hybrid halide perovskites cannot be used for the same application because of their relatively high electronic conductivity and loss. However, they are suitable for an electrode material or an electrolyte in supercapacitors.[254-256]

< **Figure 10**>

**Figure 10:** A schematic of energy density storage estimation (highlighted shaded area) from the Polarization–Electric field *(P-E)* behavior of (a) ferroelectric, (b) relaxor ferroelectric and (c) antiferroelectric materials. Adapted from references[246, 251, 253]

### 7.2. Hybrid Halide Perovskites for Supercapacitors

Recently, Zhou reported a thin film electrochemical capacitor based on an organo-lead-triiodide perovskite which exhibited a stable capacitance beyond $10^4$ cycles[256]. Such an approach provides a novel dimension for studying the ionic properties of hybrid halide perovskites and developing new devices. Shortly after, a perspective article by Snaith *et al.* postulated that the high surface area and ionic mobility of hybrid halide perovskites could make it a good alternative as an electrode or electrolyte in a supercapacitor. [254, 255] The major difference between a supercapacitor and a parallel plate capacitor is that the dielectric layer is replaced by an electrolyte in-between the electrodes. The ions in the electrolyte respond to the electric field, in contrast to the dipoles in a dielectric. Although research in this area has only just begun (in 2016) the presence of ferroelectric domain walls and diploes could enhance the performance of these capacitors. The same reasons could also be helpful in utilizing these materials in battery applications, which will now be described.

### 7.3. Hybrid Halide Perovskites for Batteries

In 2015, Xia *et. al* reported that $MAPbX_3$ (X=Br, I) is a potential anode material for lithium ion batteries with a storage capacity of ~330 mAh $g^{-1}$.[257] However, it is still not clear if the Li-ion was stored by intercalation or if it was a surface phenomenon. Moreover, rapid deterioration of the electrode was also a major obstruction in commercializing these materials in a battery. Further investigations in this direction revealed that post-poling ion migration is much faster at grain boundaries than within the grains[258], which supports the claim of the Xia *et. al.*[257] and has motivated researchers to develop novel perovskites to address the key challenges of achieving improved storage capacity and electrode stability at comparatively low cost, as compared to current commercially available batteries. Generally, solid state batteries are comprised of two lithium storing electrodes and an ion-conducting electrolyte. During charging, the lithium ions are driven into the anode by intercalation and the positive charge is compensated by the electrode.[259] During discharge, the Li-ions move back to the cathode and the current produced by the corresponding reverse flow of electrons is used to power the device. To make this technology viable, critical milestone must be attained, which include fast ion migration with stable capacity and long cycling life.

## 8. Materials Selection and Future Prospects

The above discussion suggests that each technology and mechanism has its own desired characteristics, efficiency limit and corresponding advantages and disadvantages, such as a high band gap and low absorption coefficient restricts the photovoltaic performance in ferroelectrics, while technological constraints restrict single junction and silicon photovoltaics to 30% (Shockley and Queisser limit)[260]. Similarly, the ability to achieve a high thermoelectric operating temperature for ferroelectrics is an issue, while the absence of the desired electrical conductivity in hybrid perovskites is an additional barrier for thermoelectric applications. Consequently, there is a need to develop new materials with a suitable range of properties for effective utilization of available solar and thermal resources. In parallel, the approach of simultaneously engaging multiple energy conversion mechanisms could be envisaged[19, 261, 262]. One such illustration is reported by Zhang *et.al.*[160] who enlarged the photovoltaic response of a device by combining the classical photovoltaic effect with the ferroelectric photovoltaic effect; this was achieved by sandwiching lanthanum-modified lead zirconate titanate between two low work function metal electrodes. Using a similar concept, Zhu *et. al.* took advantage of a *piezo-phototronic* effect by using a ZnO nanowire array on a silicon substrate to achieve enhanced efficiency[263]. Other examples include the fabrication of a multi-functional nano-generator using $PbZr_xTi_{1-x}O_3$ that was integrated with an air-driven nylon membrane and thermoelectric module[264], and the hybridization of electrical nano-generators and solar cells with supercapacitors for self-powered wearable electronic textiles[265]. Park *et. al.* demonstrated a hybrid energy conversion system with integrated pyroelectric and thermoelectric modules to harvest solar energy across the full spectral range and demonstrated switching of electrochromic displays using the approach. In an analogous approach, Kim *et. al.* demonstrated enhanced energy collection by integrating a photothermal, pyroelectric and thermoelectric module with a solar cell.[266] A detailed state-of-the-art summary of nano-generator technologies based on mechanical (piezoelectric[267] and triboelectric[268]), thermal (pyroelectric[269] and thermoelectric[270]) and solar energy harvesting, their coupled mechanisms to harvest energy from multiple sources[271-275] and the possibilities of integrating energy harvesting devices with storage units[276, 277] can be found in the literature[261, 262]. However, these systems require distinct materials and, therefore, a more complex device structure is required, compared to a device operating on a single material, which is likely to increase losses. In contrast, a single material with an optimal combination of desired characteristics could enhance research effort in this direction.

To some extent most of the material property requirements for each application overlap with one another. Therefore, materials with overlapping desired properties for different applications could be used for a multiple energy conversion system. This will not only aid engineers and physicists in selecting an appropriate material from the existing pool of materials, but also provide insight for chemists and materials scientists for developing new materials. This could be achieved by employing appropriate materials selection techniques [278-284] or a detailed understanding of the individual material requirements which are discussed in the following sub-sections Similar approaches be used to tune materials for a multiple energy conversion system.

**8.1. Material Requirements for Hybrid Perovskite Solar Cells**

Initially, a suitable material for photovoltaic applications must have good light absorption, carrier generation, carrier lifetime and mobility [285-287]. The material should have a high absorption coefficient, which governs the generation of free carriers or excitons subjected to the binding energy, temperature and carrier density[287]. A perovskite layer should have modest mobility[288] and a sufficiently high diffusion length which is a measurement of carrier transportation to the electrode before recombination[285]. In order to build efficient devices it is essential to optimize the film thickness and develop deposition and film treatment techniques for reliably producing good quality films[287]. It has been suggested that an appropriate selection of cations could facilitate spontaneous electrical polarization in these materials[289]. This will result in the creation of internal junctions at ferroelectric domains and help in the separation of photo-excited charge carriers. Therefore, an improved knowledge of the fundamental ferroelectric nature of potential materials could significantly help in designing smart materials with better performance[290]. In addition, the presence of ferroelectric domains could aid the reduction of segregation assisted recombination of charge carriers[285]. The transportation could be adversely affected by the presence of defects, which is often increased by doping since it enhances carrier scattering and thus influences conductivity, minority carrier lifetime and mobility. Doping not only affects electrical transport, but also governs the operating mechanism of the device[287]. The attractive photovoltaic performance of hybrid halide perovskites is often attributed to a low density of trap defects [288, 291-293] or relatively shallow traps[294-296]. In addition, interface engineering that includes the development of alternate electron and hole transport materials or the use of doping can further improve the interface (morphologically and electrically) for charge transport[285]. Since the charge transport capability of a material also governs its suitability for supercapacitor and battery applications, any improvement to achieve better transport dynamics will also increase the suitability of these materials for supercapacitor and battery applications.

## 8.2. Material Requirements for Ferroelectric Photovoltaics

As with the hybrid halide perovskites, the photovoltaic performance of ferroelectrics is also governed by light absorption, generation and separation of excitons, and the transportation dynamics of the charge carriers. Unlike the hybrid halide perovskites, ferroelectrics typically have a high band gap and low absorption coefficients. In this context, several attempts have been made to reduce their band gap and to enhance absorption using doping[297, 298], alloying[187] and oxygen vacancies[163, 299-301]. Though the presence of oxygen vacancies will enhance the conductivity of the ferroelectric and aid in charge transportation but a significant increase in conductivity will also lead to diminished ferroelectric polarization and enhanced leakage currents. Therefore, it is important to develop a trade-off between conductivity and polarization. In addition, the internal photoelectric effect can be used to enhance light absorption[160]. Although the absorption coefficient of ferroelectrics is typically low, their charge dissociation efficiency, due to a low binding energy, is high in contrast to hybrid halide perovskites. In general, ferroelectrics possess a high dielectric constant; which is inversely proportional to the binding energy of the excitons. However, the BPVE in ferroelectrics is governed by the non-centrosymmetric potential which is further dependent on the direction of polarization[302, 303]. The pre-decay shift in excited charge carriers is merely a few angstrom in the direction of polarization which results in small photocurrents.[46] Therefore, it becomes important to increase the non-centrosymmetry in ferroelectrics.[136, 196, 304, 305] The idea of enhancing the photocurrent by increasing the non-centrosymmetry is also supported by experiments.[306-308] For the final stage of charge collection, a reduction in the ferroelectric film thickness is beneficial, but the photovoltages are decreased.[132] Another method for increasing the collecting charge electric field is by forming an extending depletion using metal/intrinsic semiconductor/metal structures.[309] This could also be achieved by replacing the metal electrode with semiconductors which increases the depolarization field and charge collection efficiency by lowering the screening of spontaneous polarization.[121, 150] The depolarization field is strongly dependent on the screening conditions at the interface and the film thickness.[131, 146, 148, 149] Therefore, interface engineering and film thickness optimization are essential for high photovoltaic responses of ferroelectrics.

### 8.3. Material Requirements for Thermo-electrical energy conversion and Storage

The energy conversion using pyroelectric and thermoelectric effects could be termed as thermoelectrical energy conversion. For high pyroelectric energy conversion, the material should have a high pyroelectric coefficient to have a large change in polarization with respect to temperature fluctuation *(∂P/∂T)*. Consequently, dielectric anomalies, phase transitions or instantaneous switching, creation/destruction of crystal domains and the Curie temperature should lie within the operating temperature range. In addition, the materials should possess a high breakdown strength and high dielectric constant to have a large change in polarization with a variation in applied electric field. Importantly, it should exhibit low losses and leakage currents[18, 223, 310 311]. Interestingly, the requirements for high pyroelectric energy conversion are similar to the requirements for high energy storage in ferroelectric capacitors[16]. As far as the requirements for thermoelectric energy conversions are concerned, ferroelectrics lie far behind the hybrid halide perovskites. However, to date no experimental study has confirmed the thermoelectric performance of the hybrid halide perovskites. Only theoretical calculations have been performed and suggestions have been made to tune the electrical conductivity. However, one report claims that using a $CaMnO_3$ ferroelectric buffer layer with a hybrid halide perovskite could help in increasing the open circuit voltage and provide a better thermoelectric output.[312] Similar attempts could help in creation of efficient energy conversion devices and makes it interesting to compile the materials requirements for possible multiple energy conversion devices.

## 9. Analogy in Material Requirements and Conclusions

There are theoretical reports where the pyroelectric effect is found to support the photovoltaic performance of the material.[313-315] Recently, the possibility of tuning piezoelectric[316], pyroelectric and dielectric properties simply by illumination has been illustrated.[317, 318] Moreover, it is also possible to take advantage of piezoelectricity to enhance pyroelectric[19, 26], solar cell performance[19] and the performance of self-powered photodetectors[319]. Interestingly, the ideal material requirements for photovoltaics are the same as for supercapacitor and battery applications. The same requirements with a tuned electrical conductivity also makes a hybrid halide perovskite suitable for thermoelectric applications. In this context, the efforts of doping and creating an understanding of the effect of oxygen vacancies could be adopted from ferroelectrics. Importantly, ideal properties for pyroelectric energy conversion and ferroelectric capacitors supplement photovoltaic effects in both ferroelectrics and perovskites. A stable hybrid organic-inorganic flexible ferroelectric material with a low band gap, high absorption coefficient and better charge transportation dynamics could become an excellent material for photovoltaic, supercapacitor and battery applications while a similar material with low thermal conductivity will be beneficial for thermoelectric effects. For pyroelectric energy conversion and ferroelectric capacitors, the leakage currents should be low and the existing properties with a high band gap, break-down strength and dielectric constant with low conductivity in a hybrid organic-inorganic material could provide a better conversion efficiency and storage. Such perovskite materials with improved storage and conversion efficiencies could be integrated together for better utilization of available thermal and solar resource. Clearly, if multiple energy conversion mechanisms are anticipated there is a need to develop synergies in the multifunctional properties of these materials. In summary, the information compiled herewith is aimed at motivating researchers to realize the potential of existing perovskites and to develop novel hybrid halide ferroelectric perovskites for multiple energy conversion and storage systems.

**Acknowledgements**

Richa acknowledge support from the Center of Excellence in Nanoelectronics (CEN), IIT Bombay, India. Gaurav, Jae, Anita and Jan acknowledge support by the Australian Government via the Australian Research Council (ARC) through Discovery Grants and the Australian Renewable Energy Agency.

**Table 1:** Photovoltaic performance of ferroelectric materials.

| **Material** | **Fabrication Method** | **Photovoltage** | | **Photocurrent** | | | **Efficiency $P_{out}/P_{in}$ (%)** | **Working Mechanism** |
|---|---|---|---|---|---|---|---|---|
| | | **$V_{oc}$ (V)** | **L (µm)** | **$I_{sc}$ (µA $cm^{-2}$)** | **Light Intensity (mW $cm^{-2}$)** | **Light Wavelength (nm)** | | |
| **Pt/PLZT(3/52/48)/ITO**[132] | MOD | 0.86 | 4 | 1700 | 150 | - | - | **BPVE** |
| | | ~496 | 2400 | ~16.8 | | | | |
| **$BaTiO_3$**[47] | Sputtering, FIB, PLD | 8 | - | 17 | 100-470 | 405 | 100(EQE) | **BPVE** |
| **Pt/$Bi_2FECrO_6$/Nb–$SrTiO_3$**[320] | PLD, Sputtering | 0.74 | 0.125 | 990 | 1.5 | 635 | 6.5 | **BPVE** |
| **Au/PLWZT(3/52/48)/Au**[165] | Solution Coating | 7.0 | 25 | - | 1.11 | 365 | - | **BPVE** |
| **ITO/PZT(53/47)/ITO**[151] | PLD | 0.45 | 0.4 | 0.006 | 0.45 | - | 0.6 | **SCE & BPVE** |
| **Fe/BFO/LSM/$SrTiO_3$**[142] | PLD, Epitaxial | 0.21 | - | 48 | 20 | W-light | - | **SCE & BPVE** |
| **Mg/PLZT(3/53/48)/ITO**[160] | HPC | 8.34 | 300 | 3.25 | 100 | Sunlight | - | **PE & BPVE** |
| **Pt:Pd/BFO/Pt:Pd**[116] | Mix-flux Technique | 6 - 30 | 50-300 | $10^7$-$10^8$ | 40,000 | 405 | 40(IQE) | **TE & BPVE** |
| **Pt/PZT(20/80)/Pt**[173] | Sputtering | - | 0.36 | ~8 | 10 | 350-450 | - | **SCE & BPVE** |
| **Au/PLWZT(3/52/48)/Au**[162] | Sol-gel | 0.6 | 0.706 | - | 0.74 | 365 | - | **SCE** |
| **Pt/PZT(52/48)/Pt or Ni**[321] | Sol-gel | ~0.8 | 0.2 | ~0.03 | 0.05 | 300-390 | - | **SCE** |
| **$SrRuO_3$/BFO/ ITO** | MOCVD | 0.8 - 0.9 | 0.2 | 1500 | 285 | Sunlight | 10 (EQE) | **SCE** |
| **Au/BFO/Au[322]** | Mix-flux Technique | ~0.08 | 80 | 8.219 | <20 | 532 | - | **SCE** |
| **$SrRuO_3$/BFO/ Au[323]** | Sputtering, Epitaxial | 0.286 | 0.17 | 0.4 | 750 | 435 | - | **SCE** |
| **Au/BFO/Au[163]** | Mix-flux Technique | ~0.7 | 60 | 1.58 | 20 | 532 | 1.5 (EQE) | **SCE** |
| **Nb-doped $SrTiO_3$/BFO/ Au[194]** | PLD | ~0.15 | 0.1 | 6000 | 285 | W-light | 0.03 | **SCE** |
| **ITO/PZT/$Cu_2O$/Pt[324]** | Sol-gel | 0.6 | 270 | 4800 | 100 | Sunlight | 0.57 | **SCE** |
| **Pt/Poly-BFO/Au & ITO[150]** | Sol-gel | 0.1 | 0.3 | ~1 | 450 | 340 | - | **SCE & DF** |
| **Graphene/Poly-BFO/Pt[309]** | Sol-gel | 0.20 | 0.3 | 2800 | 100 | Sunlight | - | **MIM-SCE** |
| **Nb: $SrTiO_3$/PLZT (3/52/48)/LSM[121]** | Sputtering, Epitaxial | ~0.7 | 0.068 | ~0.8 | 0.059 | - | 0.28 | **DF** |
| **Pt/BFO/Pt[41]** | MOCVD | 16 | 200 | 120 | 285 | W-light | $10^{-3}$ (EQE) | **DW** |
| **Pt/BFO/Pt[40]** | MOCVD | 0.014 | One DW | 50 | 100 | W-light | 10(IQE) | **DW** |
| **FTO/Poly-BFO/AZO[152]** | CSD | 0.63 | - | 130 | 100 | Sunlight | 7 (EQE) | **BI & DF** |
| **ZnO : Al/BFO/LSC[325]** | PLD | 0.22 | 0.35 | ~5 | 1 | W-light | - | **-** |
| **Ag/Pr-doped BFO NTs/Ag[196]** | Chemical Technique | 0.21 | - | - | 10 | Sunlight | ~0.5 | **-** |

BPVE=bulk photovoltaic effect; DF=depolarization field effect; SCE=Schottky contact effect; DW=domain wall effect; MIM=metal/insulator/metal junction, PE=photoelectric effect;

BI=built-in potential due to asymmetric electrodes; TE=tip enhancement effect; MOD=metal–organic decomposition; MOVCD=metal–organic vapor phase epitaxial; FIB=Focused ion beam milling; PLD=pulsed laser deposition; HPC=hot-pressing calcinations; CSD=chemical solution deposition; IQE=internal quantum efficiency; EQE= external quantum efficiency

**Table 2:** Comparison of energy density and corresponding conditions for selected compositions.

| Material | $T_{Low}$ (K) | $T_{High}$ (K) | $E_{Low}$ ($MVm^{-1}$) | $E_{High}$ ($MVm^{-1}$) | Energy Density ($kJm^{-3}cycle^{-1}$) |
|---|---|---|---|---|---|
| **73/27 P(VDF–TrFE)^** [217] | 296 | 340 | 23 | 53 | 30 |
| **PZN-4.5PT**[37] | 373 | 433 | 0 | 2.0 | 217 |
| **PZN-5.5PT**[35] | 373 | 463 | 0 | 1.2 | 150 |
| **PMN-10PT**[35] | 303 | 353 | 0 | 3.5 | 186 |
| **PMN-32PT**[23] | 353 | 443 | 0 | 0.9 | 100 |
| **60/40 P(VDF–TrFE)^** [199] | 331 | 350 | 4.1 | 47.2 | 52 |
| **PNZST**[216] | 418 | 448 | 0.8 | 3.2 | 300 |
| **8/65/35 PLZT**[#24] | 298 | 433 | 0.2 | 7.5 | 888 |
| **BNT–ST–BLT**[14] | 293 | 413 | 0.1 | 6 | 2130 |
| **KNTM**[15] | 413 | 433 | 0.15 | 0.15 | 629 |
| **BNLT**[13] | 298 | 393 | 0.1 | 11.2 | 1146 |
| **BNKT**[13] | 298 | 383 | 0.1 | 5.2 | 1986 |
| **BNK-BST**[17] | 293 | 433 | 0.1 | 4.0 | 1523 |
| **PLZST (x=0.2)** [#326] | 293 | 493 | 30 | 40 | 6800 |
| **YBFO^**[16] | 15 | 300 | 0.1 | 4 | 7570 |
| **PLZST (x=0.18)**[# 327] | 298 | 573 | 30 | 90 | 7800 |
| **0.67PMN-0.33PT^**[223] | 303 | 323 | 0 | 60 | 6500 |
| **0.68PMN-0.32PT^**[223] | 303 | 323 | 0 | 60 | 8000 |
| **$Hf_{0.2}Zr_{0.8}O_2$**[328, 329] | 273 | 423 | 0 | 326 | 11549 |
| **PZT/CFO/PZT^**[20] | 100 | 300 | 0 | 40 | 47372 |

#Thick Films; ^Thin Films

***Figures:***

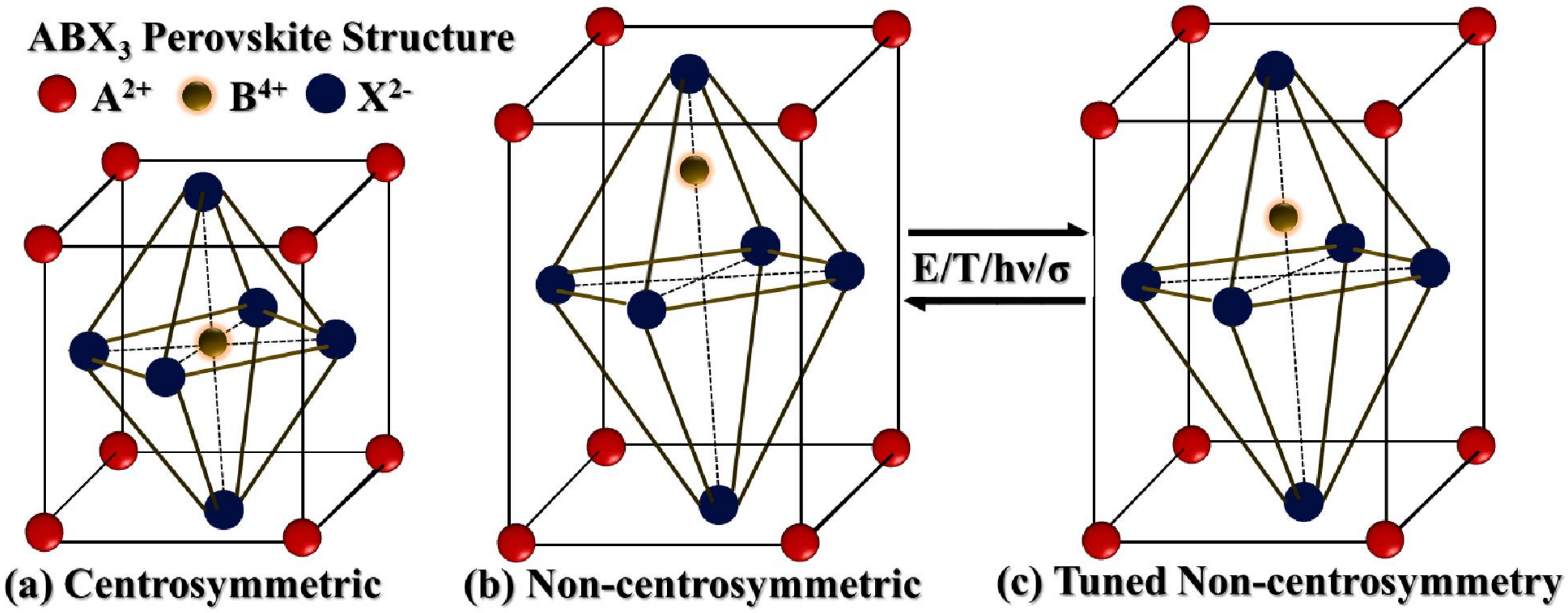

**Figure 1**

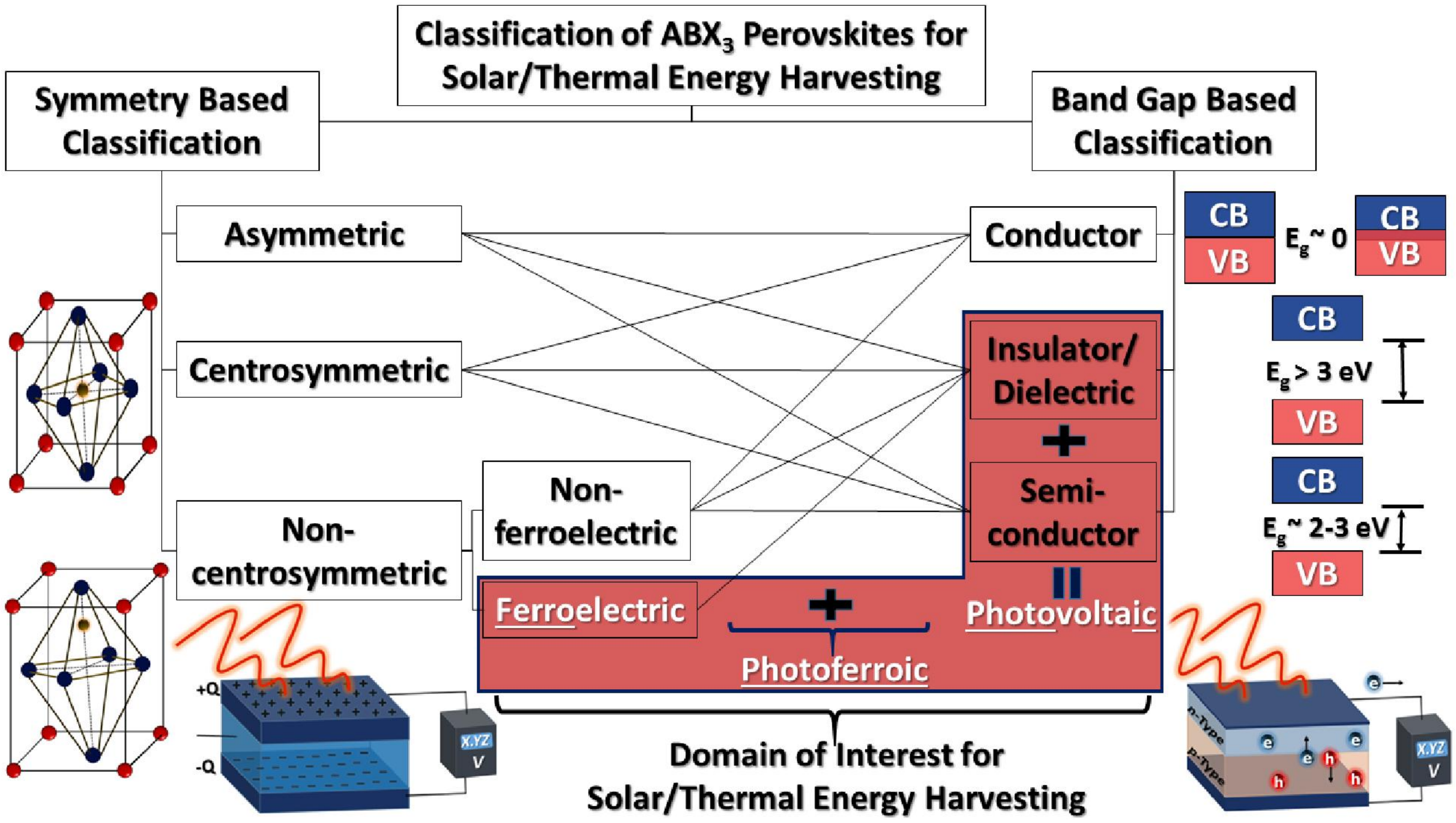

**Figure 2**

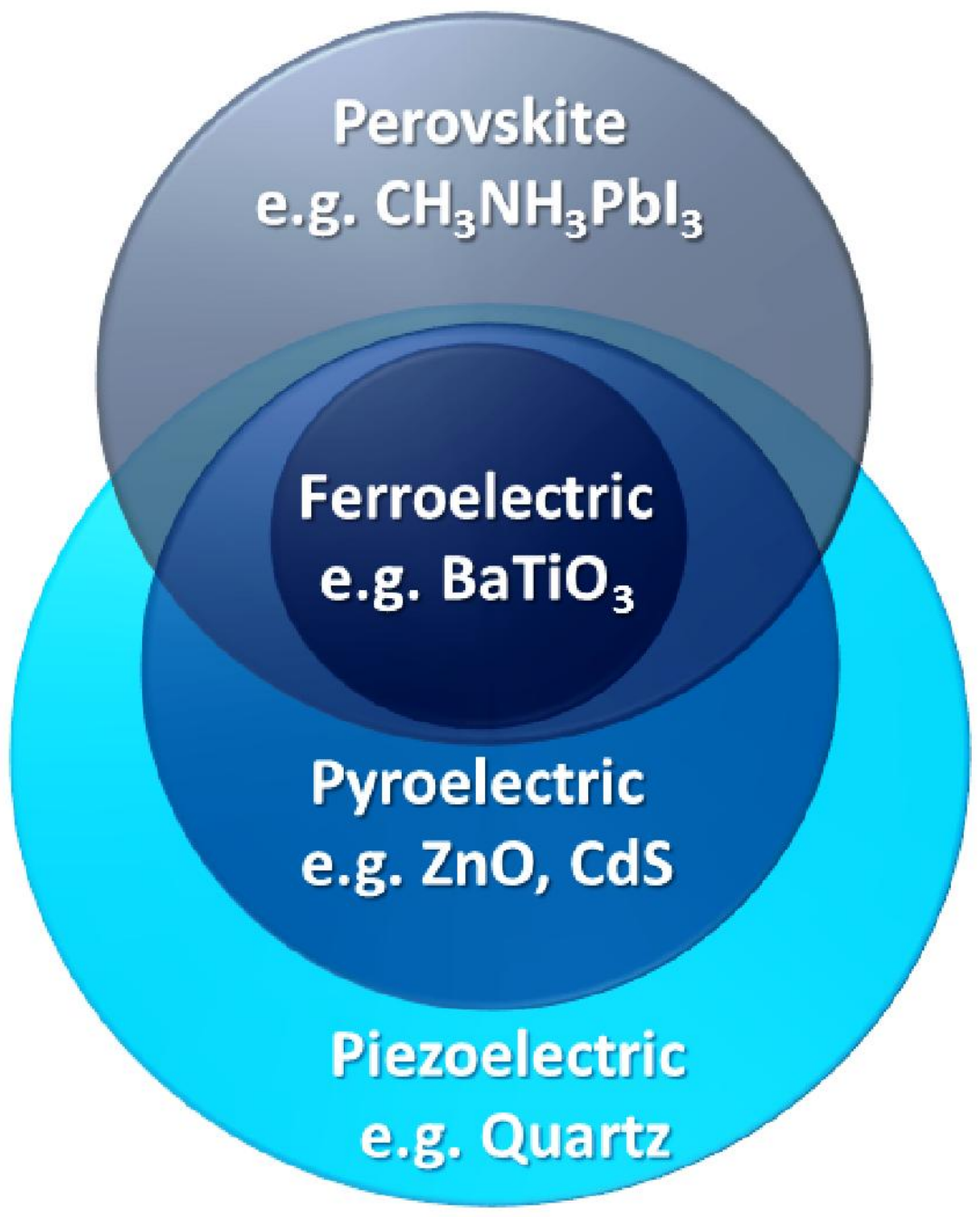

**Figure 3**

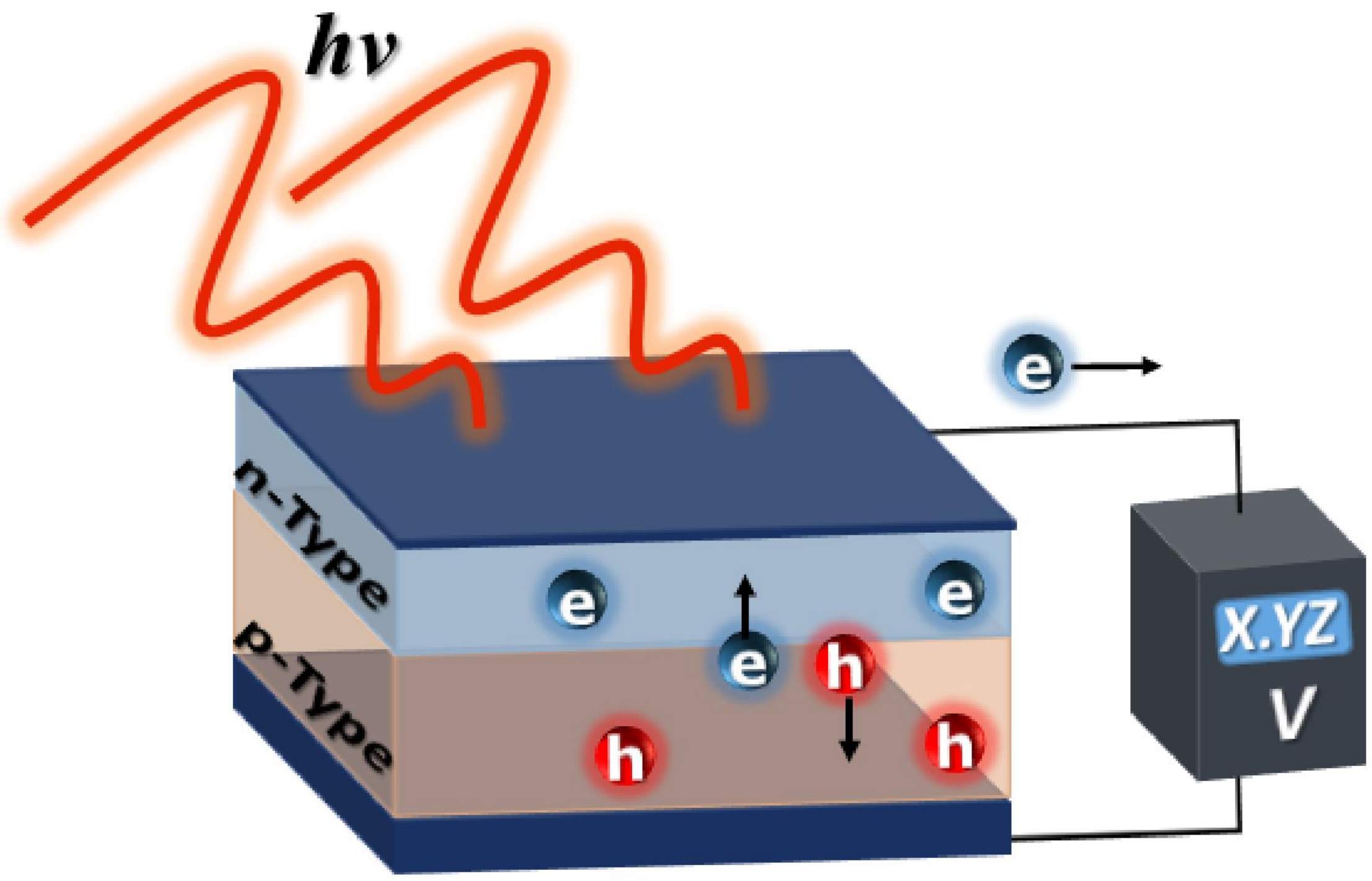

**Figure 4**

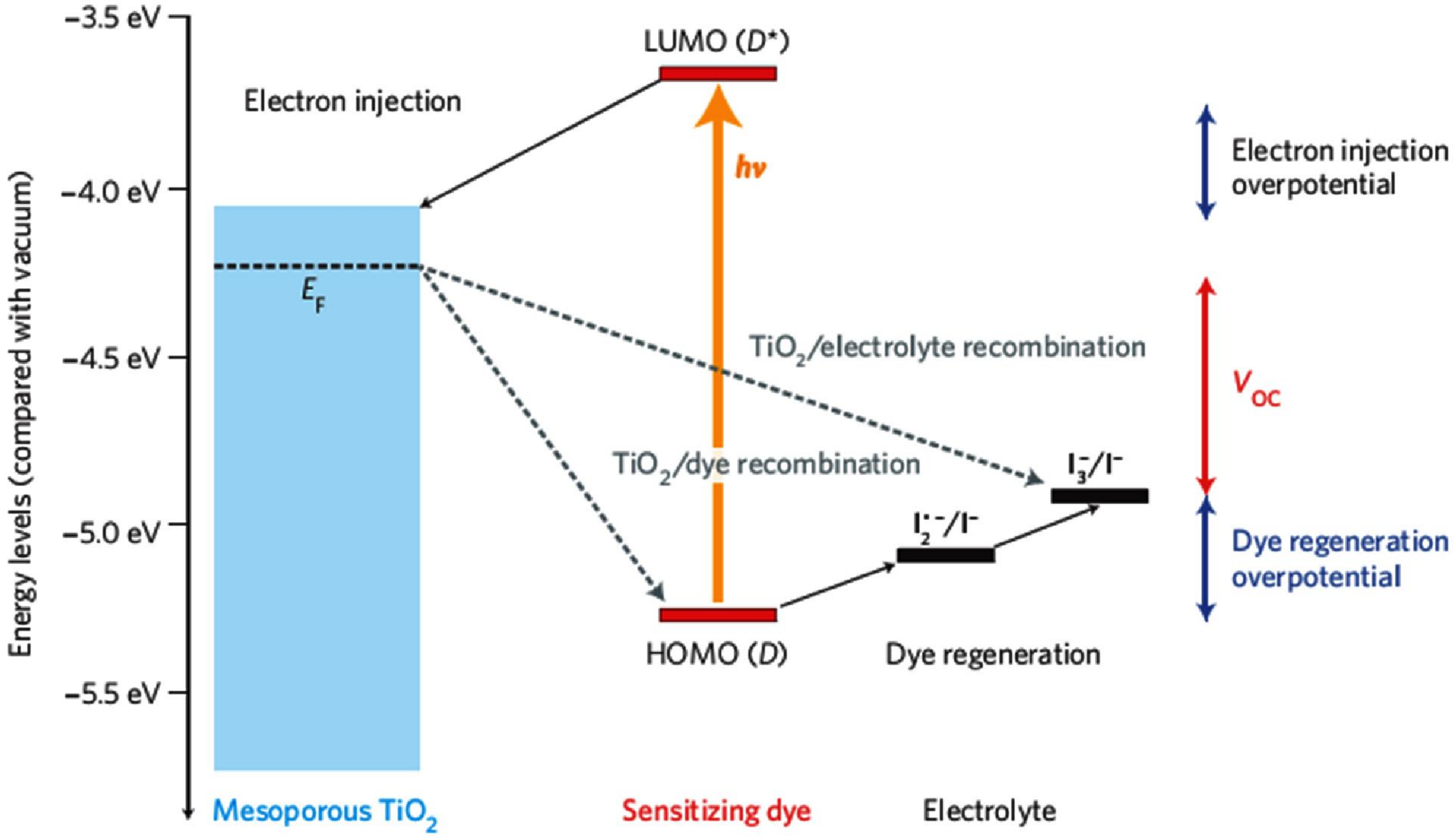

**Figure 5**

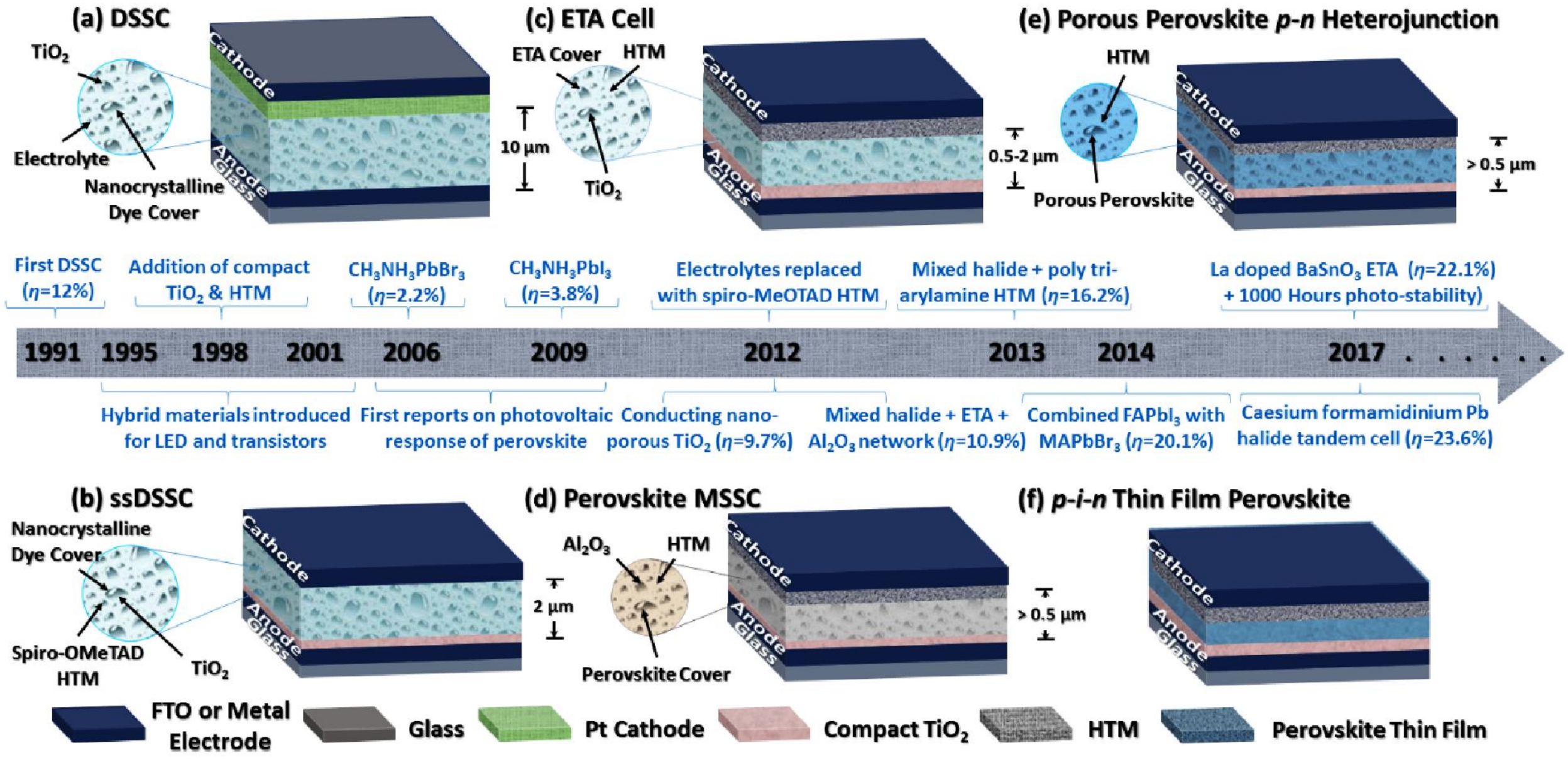

**Figure 6**

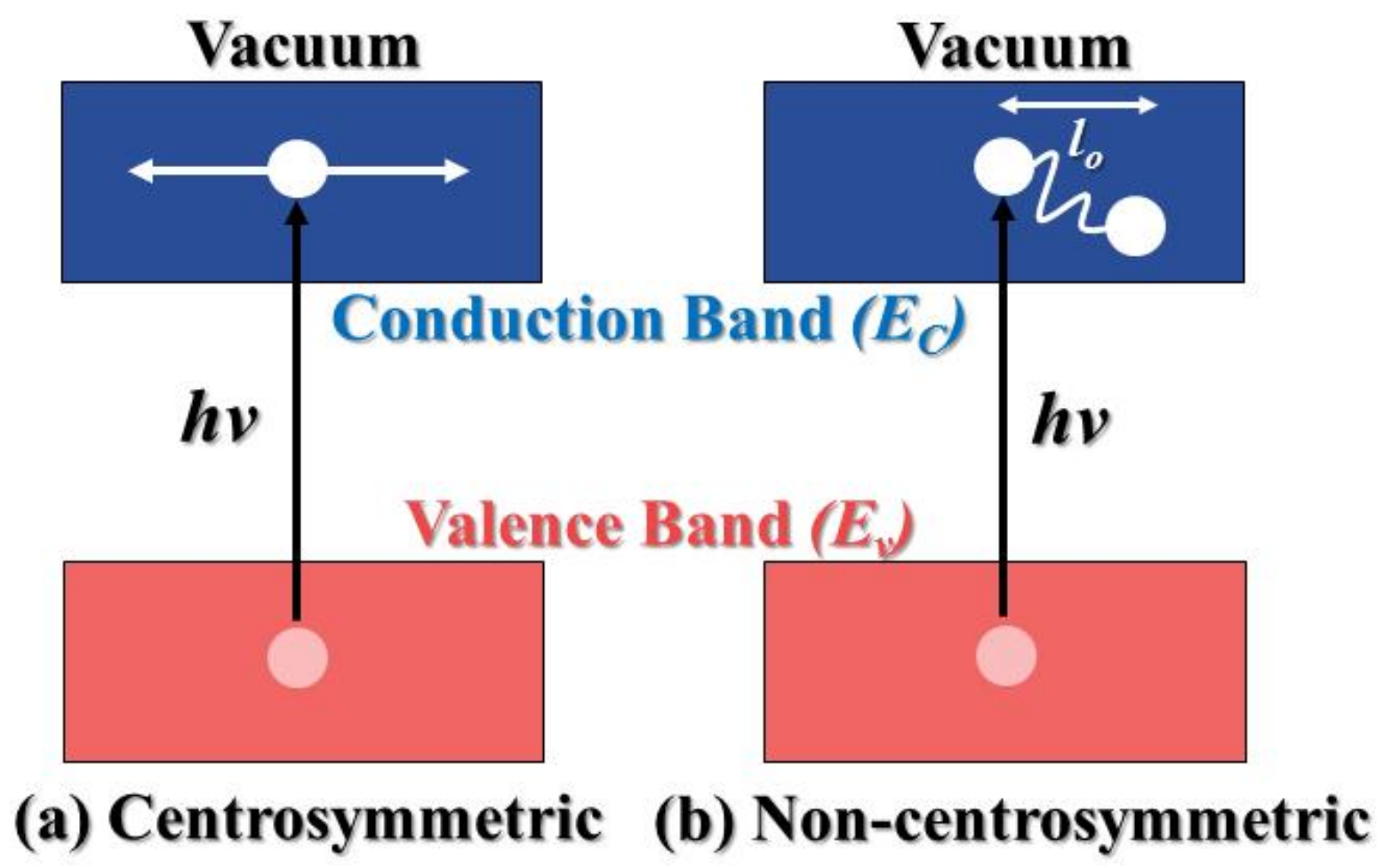

Figure 7

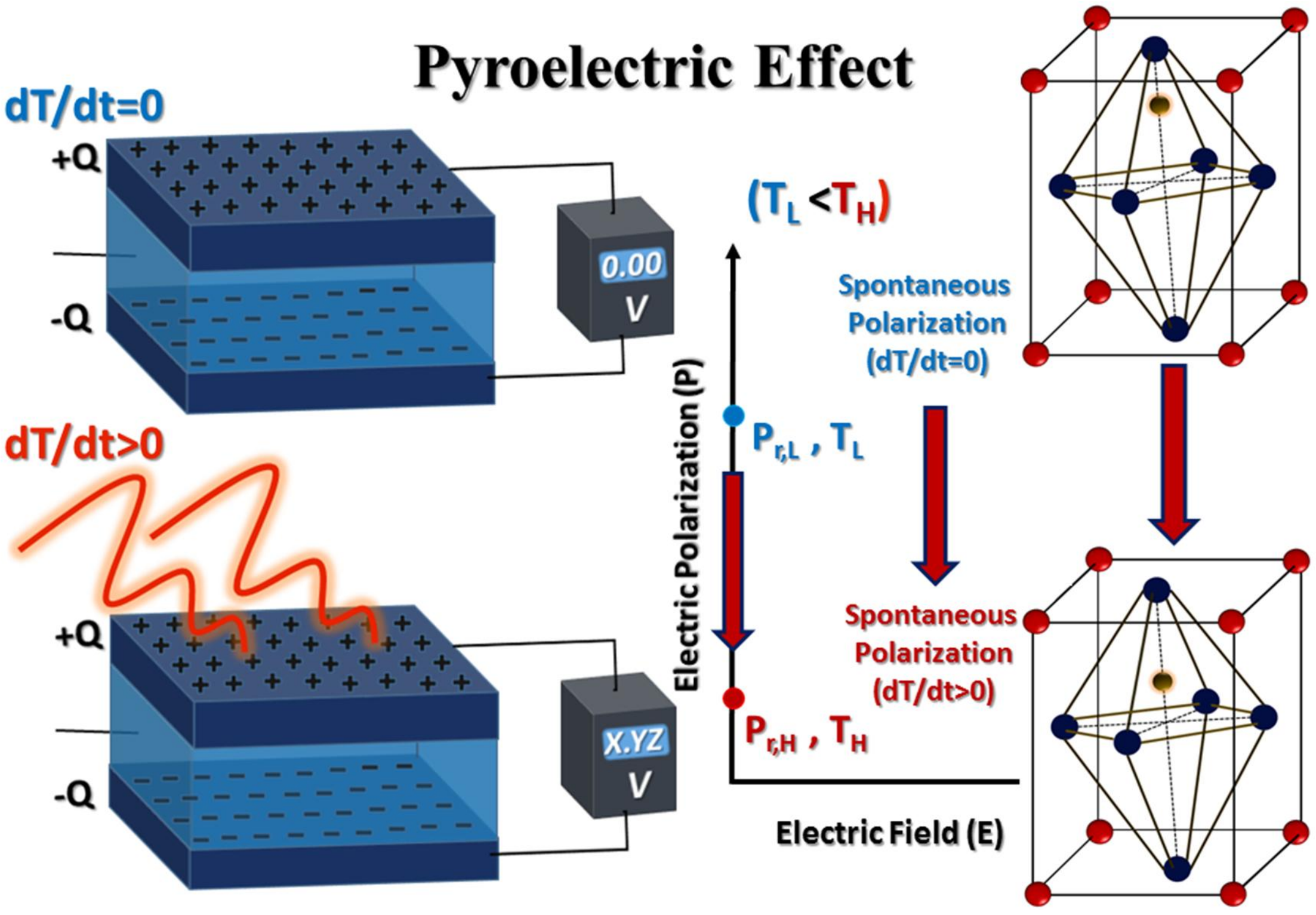

Figure 8

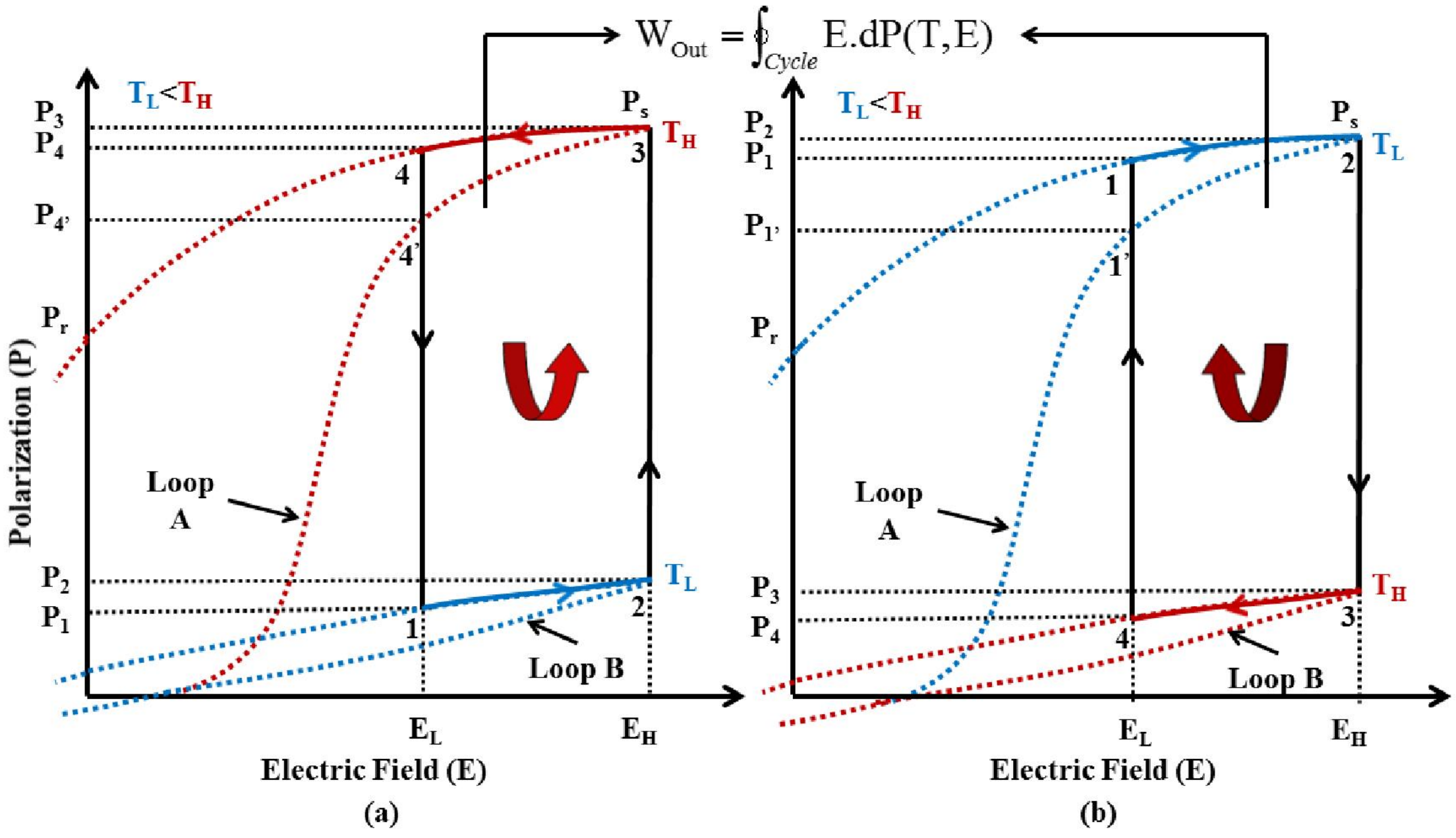

**Figure 9**

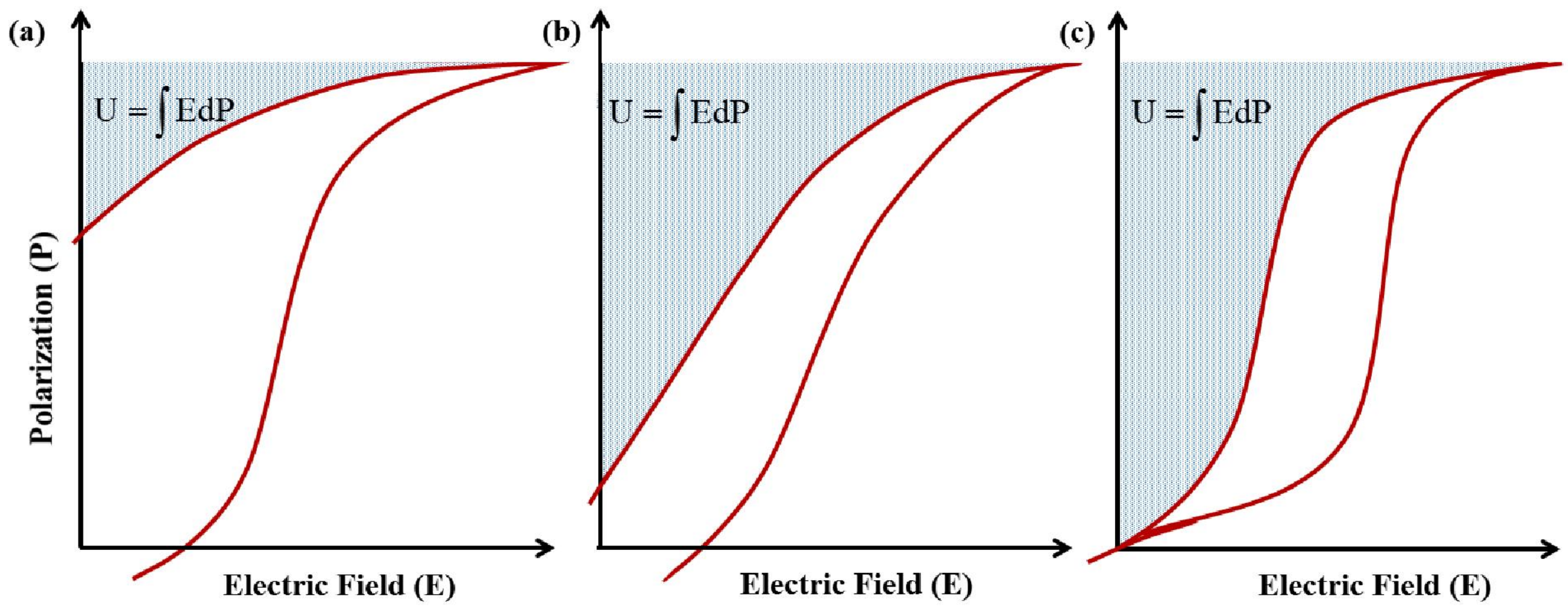

**Figure 10**